\pdfoutput=1
\documentclass[10pt,conference]{IEEEtran}
\usepackage[T1]{fontenc}
\usepackage{microtype}
\usepackage{cite}
\usepackage{xurl}
\usepackage[hidelinks]{hyperref}
\title{Disposable Implementations:\\ What Is Software Engineering For?}
\author{\IEEEauthorblockN{Christoph Treude}
\IEEEauthorblockA{Singapore Management University\\Singapore}}
\begin{document}
\maketitle
\begin{abstract}
Existing implementations are usually maintained for most of their life and replaced only as an exception. Coding agents may change that. If agents can produce new implementations from requirements, tests, architectural constraints, and operational records that survive replacement, changing a system may no longer require modifying its existing codebase. We call this process \emph{regeneration}. An implementation is \emph{disposable} when regeneration becomes a routine alternative to maintenance. Regenerating and verifying a clean implementation from current requirements and tests may be cheaper than having agents navigate (and potentially compound) existing technical debt, and could eliminate much of that debt. Yet existing code may contain knowledge absent from the preserved artifacts, including safeguards introduced after failures and assumptions that were never documented. A regenerated implementation may pass the available tests and still lose behavior that matters in production. This leaves open how to establish continuity across replacements and whether regeneration reduces maintenance effort or shifts it to the artifacts and processes that support it. We examine these questions through an exploratory review of practitioner literature. Published examples show that agents can replace existing implementations, but do not establish that repeated regeneration is safe or economical. If implementations become disposable, what should software engineering preserve?
\end{abstract}

\section{The Idea}\label{sec:idea}
Lehman's first law of software evolution, \emph{continuing change}, reads: ``A program that is used and that as an implementation of its specification reflects some other reality, undergoes continual change or becomes progressively less useful. The change or decay process continues until it is judged more cost effective to replace the system with a recreated version''~\cite{lehman1980}. The cost of writing a replacement has fallen considerably since Lehman formulated his first law. Rather than modifying existing code, coding agents could produce new implementations from requirements, interface contracts, tests, architectural constraints, and operational records preserved across replacements. We call this process \emph{regeneration}. An implementation is \emph{disposable} when regeneration becomes a routine alternative to maintenance rather than a last resort at the end of a codebase's life. The system persists; the implementation does not.

An existing implementation embodies accumulated effort and knowledge, but it also carries technical debt. Coding agents reduce the effort needed to write a replacement, though not necessarily the cost of recovering the knowledge. As Chad Fowler puts it, ``When code is cheap to generate, preserving it at all costs becomes irrational''~\cite{fowler2025}. Rather than having agents navigate years of accumulated technical debt to make a change (and potentially compound that debt), it may be cheaper to regenerate and verify a clean implementation from current requirements and tests, potentially eliminating much of that debt in the process.

Whether replacement makes sense depends on the software. Kanat-Alexander places software on a spectrum from \emph{disposable} to \emph{durable} according to four factors: the consequences of failure, the number of people and systems that depend on the software, the cost of rewriting it from scratch, and the cost of maintaining it~\cite{kanatalexander2026}. At one end are applications that can be discarded and rewritten with little consequence; at the other are systems whose failure would be costly and on which other systems depend. He argues that rigorous software engineering remains necessary for durable systems, regardless of who writes the code. Coding agents primarily reduce one of his four factors: the cost of writing a replacement. They do not necessarily reduce the cost of verifying the replacement, integrating it with dependent systems, or migrating existing state. Kanat-Alexander also suggests making software more disposable by limiting the consequences of failure or the number of systems that depend on it, for example by restricting access to customer data. We instead ask whether an implementation can become disposable while the system it supports remains durable.

Suppose AI agents maintain a service that other systems depend on. A new requirement changes how the service handles failed transactions. Rather than modify the code, the agents generate a replacement from the artifacts preserved across replacements. It passes the acceptance checks and is deployed, although its structure looks little like the previous version. The old code may be kept for reference, but it is no longer the starting point for the next change. Now suppose the old code contained a retry safeguard, added after a production incident. Nobody had documented or tested it. In Chad Fowler's phrase, the implementation remembered what the organization had forgotten~\cite{fowler2026}. The agents leave out the safeguard because none of the preserved artifacts describes it, and the replacement loses a lesson learned in operation. \emph{What must carry over between implementations for us to regard them as versions of the same system?}

Software engineering may need to focus less on preserving an implementation and more on producing new implementations and establishing that they are acceptable. Three things that have usually been maintained together may come apart: the durability of the system, the continuity of its implementation, and the knowledge required to change it safely. We develop this argument through an exploratory review of practitioner literature.

\section{Why It Matters Now}
Planned replacement and generation from longer-lived artifacts both predate coding agents. In 2014, Martin Fowler argued that ``often the best code you can write now is code you'll discard in a couple of years time''~\cite{mfowler2014}; his \emph{sacrificial architecture} allows an implementation that fits one stage of a system's life to be replaced at another. Model-driven engineering made models the principal asset and used defined languages and transformations to generate implementations~\cite{schmidt2006}. Coding agents differ in that they can produce the replacement themselves from incomplete natural-language requirements, existing code and documents, tool feedback, and executable checks. B\"ockeler distinguishes \emph{spec-first}, \emph{spec-anchored}, and \emph{spec-as-source} workflows, and questions whether the latter two might inherit the inflexibility of model-driven development as well as the non-determinism of language models~\cite{boeckeler2026}.

Chad Fowler describes \emph{regenerative software}, arguing that the system, rather than its codebase, is the asset~\cite{fowler2025}. He asks what survives ``when implementation becomes disposable''~\cite{fowler2026c}. We ask when replacing an implementation benefits the system it supports.

\section{What We Do Not Yet Understand}
\textbf{What counts as continuity?} A regenerated service is not intended to be a copy of its predecessor: a new requirement is what prompted the replacement. It may preserve its public API and pass acceptance tests while changing performance, failure behavior, security properties, or data migration guarantees. Should continuity be judged against observable behavior, stakeholder intent, persistent data, contractual obligations, or some combination? Who makes that judgment if agents produce both the implementation and its tests?

\textbf{What knowledge must persist?} Requirements, tests, architectural decisions, runbooks, and production observations could all be preserved across regenerations. They will not necessarily capture the exceptions and compromises embedded in an existing implementation. Naur argued in 1985 that the theory needed to understand and extend a program resides in the people who developed it and cannot be fully conveyed through code or documentation~\cite{naur1985}. Piccoli extends this concern to coding agents: when an agent makes fine-grained design decisions without communicating them, software ``becomes legacy immediately upon creation,'' because neither the developer nor the agent acquires the relevant understanding~\cite{piccoli2026}. Parnas identified two sources of software aging: neglecting to adapt software to changing needs, and making changes without understanding the original design~\cite{parnas1994}. Regeneration can remove structural damage left by changes, but an implementation generated without the original design concept can repeat the same mistakes. What needs to survive replacement might be a behavior, a constraint, a decision, a component boundary, or an assurance claim rather than a file or repository.

\textbf{Where does maintenance effort go?} Lehman's laws predict that complexity grows as software changes unless effort is spent controlling it~\cite{lehman1980}. Perry and Wolf describe a related decay, \emph{architectural drift}, which they attribute to ``insensitivity about the architecture'' and which makes a system harder to adapt~\cite{perry1992}. Regeneration from preserved architectural constraints might reduce such drift, but only if those constraints are complete and kept current. Agents may reduce the cost of writing code without proportionally reducing the cost of making safe changes. With regular regeneration, does complexity disappear, return in each new implementation, or shift into the artifacts and procedures used to regenerate it? Servile warns that sufficiently precise specifications ``start looking suspiciously like code''~\cite{servile2026}. Maintaining such specifications could then cost as much as maintaining the code they replace. Well-structured code may also be cheaper for agents to modify: in an informal experiment, Edwards-Alexander found that refactoring a 17,000-line agent-written Rust file reduced the input tokens needed for the same subsequent change by 83\%~\cite{edwards2026}.

\section{The Question We Would Put to the Room}
\emph{Can a software system remain durable if its implementations do not? If implementation continuity is no longer assumed, what does software engineering need to preserve, and how can we establish that the system continues to meet its obligations?}

In many systems, regeneration may remain limited to isolated components, with most implementations still worth maintaining. In other systems, continuity may depend on continuously updated evidence or on the process that generates and validates new implementations. Lulla et al. define \emph{loop engineering}, a term introduced in practitioner literature in June 2026, as the practice of designing automated control structures that repeatedly invoke coding agents~\cite{lulla2026}. For regeneration, such loops would need to preserve knowledge across implementations, verify each replacement against the system's obligations independently of the agents that produced it, and control when replacements are deployed. Continuity might then depend as much on the engineering process as on the artifacts it preserves.

\section{Why It Is Plausible}
Published examples include a large one-time replacement and the repeated regeneration of a small module. Timson reports a project that replaced 86 interdependent Go microservices, six years old and totaling roughly half a million lines of code, with a TypeScript monolith written by agents rather than humans~\cite{timson2026}. The team ran an agent-written integration test suite against both versions. Passing the same tests, however, establishes agreement only for the cases covered, not behavioral equivalence. The old code, including its bugs, served as the specification, and the first generated test suite was discarded because it checked nothing. This was a one-time migration, not repeated regeneration. Bartz-Beielstein reports an example closer to regeneration as we define it: a 239-line module was deleted from a forecasting library and rebuilt three times by Claude Opus 5.5 using records of its intent, decisions, public contract, and evaluations~\cite{bartzbeielstein2026}. All three versions passed the 37 evaluations. Those records occupied 805 lines, including 453 lines of evaluation code---3.4 times the size of the module. The experiment involved only one module.

Neither example establishes that regeneration is reliably safe or more economical than maintenance. B\"ockeler argues that code still matters when defects are difficult to detect and their consequences are serious~\cite{boeckeler2025}. Regeneration can become routine only if an engineering process can identify which behavior and constraints must survive, check that each replacement preserves them, and establish when regeneration is preferable to maintaining the existing implementation.

\section{What Would Advance It}
Laurie Voss argues that ``if you can build something that proves you got the answer, AI can get the answer much faster than you can''~\cite{voss2026}. Verifying a newly generated implementation is not enough: regeneration also requires evidence that the system continues to meet its obligations across replacements. Testing and dependability researchers could examine the risk of using the same tests to guide regeneration and approve its output. Software evolution and mining researchers could investigate how regeneration affects the interpretation of commit histories and measures of change, technical debt, and code churn. Architecture researchers might use Chad Fowler's observation that cheap code ``separates accidental irreversibility from intentional durability''~\cite{fowler2026b} to distinguish constraints worth preserving from those introduced to make a codebase easier to modify. Requirements researchers could investigate how to record and update intent so that it survives replacement, particularly when specifications are incomplete. Human--AI systems researchers could study how engineers acquire the understanding needed to supervise systems that agents repeatedly replace, and where responsibility for the result lies.

Possible evidence includes case studies of replacement projects, traces of agent-driven development, and experiments that systematically vary which artifacts are available during regeneration to determine what behavior or knowledge is lost. Repository histories alone are not enough: Lulla et al. found that repositories running agent loops rarely commit the loops' state files~\cite{lulla2026}, so studying disposability also means tracing agent executions and deployed behavior across repeated regenerations. The longer-term goal would be an empirical characterization of implementation disposability: which systems can evolve through regeneration, when it is a better option than incremental maintenance, what knowledge must survive replacement to make it viable, and where its costs and risks fall. Such a characterization needs a vocabulary for degrees of disposability and counterexamples where regeneration is unsafe or uneconomical.

Regeneration may demand greater rigor in preserving knowledge and verifying behavior across replacements, since the existing implementation can no longer be relied on to carry that knowledge forward. We need to determine which artifacts, practices, and forms of evidence are worth preserving when implementations are routinely replaced.

\textbf{Acknowledgments.} The author thanks the organizers and participants of the Thoughtworks Future of Software Development Retreat Europe (Engelberg, Switzerland, June 2026) for discussions that informed this paper.

\textbf{Use of generative AI.} ChatGPT and Claude helped refine the question, draft and critique text, verify references, and find practitioner literature. The author is responsible for the argument, citations, and final text.

\end{document}